# Making Micro- and Nano-beams by Channeling in Micro- and Nano-structures

S. Bellucci[1], V.M. Biryukov[2], Yu.A. Chesnokov[2], V. Guidi[3], W. Scandale[4]

[1]*INFN - Laboratori Nazionali di Frascati, P.O. Box 13, 00044 Frascati, Italy*

[2]*Institute for High Energy Physics, 142281 Protvino, Russia*

[3]*Department of Physics and INFN, Via Paradiso 12, I-44100 Ferrara, Italy*

[4]*CERN, Geneva, Switzerland*

## Abstract

A particle beam of very small cross-section is useful in many accelerator applications including biological and medical ones. We show the capability of the channeling technique using a micron-sized structure on a surface of a single crystal, or using a nanotube, to produce beam of a cross-section down to 1 square micrometer (or nanometer). The channeled beam can be deflected and thus well separated in angle and space from the primary and scattered particles. Monte Carlo simulation is done to evaluate the characteristics of a channeled microbeam. Emittances down to 0.1-0.001 nanometer·radian, and flux up to $10^6/\mu m^2$ per second, can be achieved for protons and ions.

**PACS codes:**

61.85.+p Channeling phenomena

41.85.-p Beam optics

87.56.-v Radiation therapy equipment

**Corresponding author:** V.M. Biryukov, Institute for High Energy Physics, Protvino, Moscow region, Ru-142281 Russia. Phone: 7-0967-713848. Fax: 7-0967-742867. E-mail: biryukov@mx.ihep.su .Web: http://beam.ihep.su/crystal.html

# 1. Introduction

Bent crystals have efficiently channeled particle beams [1] in the energy range from 3 MeV [2] to 900 GeV [3]. Today, crystals are largely used for extraction of 70-GeV protons at IHEP with efficiency reaching 85% at intensities well over $10^{12}$ particles/s, steered by silicon crystal as short as 2 mm [4]. A bent crystal (5 mm *Si*) is installed into the Yellow ring of the Relativistic Heavy Ion Collider where it channels *Au* ions and polarized protons of 100-250 GeV/u as a part of the collimation system [5].

Carbon nanotubes are cylindrical molecules made of carbon atoms. Nanotubes can be manufactured of different diameters - from a fraction of a nm to a few microns, of different lengths - from a micron up to a few millimeters, of different materials - usually carbon but also others [6]. This makes nanotubes a very interesting object for channeling research.

The purpose of the present paper is to look at how the channeling technique could be used to make beams of very small emittance. As a potential application we consider a microbeam facility being developed at BNL [7] where a variety of beams from $Fe^{+26}$ to protons of 0.1-3 GeV/u is needed with the beam size of ~10 μm at a target. A traditional approach to creation of a microbeam would be a ~20-μm-thin wire placed in a circulating beam and a set of micro-collimators cutting out a small part of the scattered-beam phase space [7]. Here the weak points can be a low flux of scattered particles in the direction of extraction line; primary and secondary particles scattered off the collimators may contaminate the microbeam.

If a channeling structure is used instead of wire, it can trap the incident particles and deliver them into a single required direction (i.e. the extraction line) instead of scattering them all the ways around. That may give a large gain in the microbeam flux. The rest of the system may be unchanged: the same set of collimators etc. Further benefit is a low divergence of the channeled beam as set by channeling acceptance; that would reduce the need in collimation down the line, and may reduce the emittance of microbeam. Finally, the channeled beam would have well-defined sharp edges and contain solely primary particles. The open point is how to make a channeling structure as small as about the size of wire, ~0.02 mm, or smaller. Below we suggest two approaches, with crystals and with nanotubes.

# 2. Crystal micro-beam

The first suggestion is to use a micron-sized structure on a surface of a single crystal; such structures are a well-developed technique [8]. The easy way to do it is to take a crystal plate, mask a strip 10 μm (or 1 μm) wide on the surface, and etch the surface to the depth of 10 (or 1) μm. That leaves a strip of 10 by 10 μm (or 1 by 1 μm) on the surface; this strip can channel particles, thus forming a microbeam. In order to separate in the angle and space the beam channeled of the strip from the particles nearby (in the crystal bulk and outside), we suggest having a strip shorter than the substrate plate (Fig.1), and bending the whole structure. That makes a perfect separation downstream.

While the size of the microbeam source is set by the strip size, the divergence in the direction of bending is set by the channeling angle, $(2E_c/pv)^{1/2}$, where $E_c$ is the critical transverse energy for channeled particles and $pv$ is the particle's momentum times velocity per unit charge. E.g., in slightly bent Si(100) with $E_c \approx 5$ eV, the 0.1-3 GeV protons have divergence of 0.05-0.2 mrad. For fully stripped ions of $Fe^{+26}$ in the range of 0.1-1 GeV/u, the divergence is 0.08-0.15 mrad. One can pick crystal channels with bigger or smaller angular acceptance.

With 1-μm source, this gives a microbeam emittance of (0.025-0.1) $\pi$ nm·radian for protons of 0.1-3 GeV, and (0.04-0.08) $\pi$ nm·radian for $Fe^{+26}$ ions of 0.1-1 GeV/u, in horizontal plane. For comparison, the horizontal emittance expected [7] from the traditional approach is 23$\pi$ nm·radian at any energy. If realized, the channeling approach would give an improvement by a factor of ≈200-1000 for protons and 300-600 for ions. It can be improved even further by collimation downstream.

In the direction orthogonal to bending, microbeam divergence equals that of the circulating beam. However, the vertical size of microbeam is set by the strip, down to ~1 μm, while in traditional approach it has to be cut by micro-collimation. Therefore, an improvement ≈100 in vertical emittance can be expected from channeling approach due to small size of the source.

## 3. Nano-beam

While crystal channeling is a well-established technique, nanotube channeling is just emerging as a beam instrument [9-13]. Here, beam can be trapped in a single nanotube cylinder of ≈1 nm diameter or in a rope consisting of many nanotubes. The depth $E_c$ of the potential well in a carbon nanotube is ~15-60 eV for channeled particles, depending on nanotube configuration [11]. The critical angle for channeling $\theta_c = (2E_c/pv)^{1/2}$ is factor of ~1.5-3 greater than with Si crystal. Provided that nanotubes can efficiently channel and deflect particle beams, they offer an interesting opportunity to make clean beams of potentially very small size, down to 1 square nanometer if needed.

We have developed a Monte Carlo code and done simulations of particle channeling in bent single-wall nanotubes, aimed at finding how useful nanotubes are for channeling of positively-charged particle beams, what kind of nanotubes are efficient for this job, and how nanotubes compare with crystals in this regard [12]. Figure 2 shows how the number of 1 GeV protons channeled through 50-μm-long nanotube depends on nanotube curvature $pv/R$ for tubes of different diameters. For comparison, also shown is the same function for Si(110) crystal (from ref.[1]). The channel length of 50 μm, with bending of 1 GeV/cm, gives the 1 GeV particles a deflection of 5 mrad - sufficient for many accelerator applications like extraction [1,3-5]. One can see from Figure 2 that a nanotube as narrow as 1 nm is comparable to silicon crystal in beam bending.

For the simulations of nanotube channeling of $Fe^{+26}$ ions and protons of 0.1-3 GeV/u, we use the tubes of 1.1-nm diameter, typical for easily manufactured carbon nanotubes. We take the curvature radius of 2 cm; then the beam energy range to be studied does nearly correspond to the *pv/R* range studied in Fig.2. We choose the nanotube bending angle of 5 mrad. Figure 3 shows two examples of the angular distribution of protons downstream of the bent nanotube, shown in the direction of bending. Similarly to pictures of bent crystal channeling, there is clear separation of channeled and nonchanneled peaks, with few particles lost (dechanneled along the tube) between them. For higher energy there is a substantial loss in efficiency due to centrifugal dechanneling, while for lower energy nearly all the particles trapped by the nanotube were channeled to its end. Overall, the transmission of particles by the tube is reasonably good on both ends of the energy range. The intermediate energies fall between the two cases shown.

The case of $Fe^{+26}$ ions is shown in Figure 4, again for both ends of the energy range of interest, 0.1 to 1.0 GeV per nucleon. Similar picture can be seen, as with protons. Overall, for the similar ratio of beam momentum per unit charge, the angular distributions of Fe ions and protons are similar. The transmission efficiency is reasonably good for all particle species. The same nanotube deflector could be used in each case, throughout the range of energies and particle species.

For 0.1-1 GeV/u ions of $Fe^{+26}$, the divergence of the channeled beam in a nanotube of arbitrary helicity like *(11,9)* is 0.24-0.77 mrad. The size of the source could be quite small. A typical nano-rope (consisting of 100-1000 nanotubes) would be a source that gives an emittance of the nano-beam of the order of 0.001 π nm·radian both horizontally and vertically, factor of 10000 down from the figure potentially achievable with a "traditional amorphous" source.

## 4. Intensity of microbeam

With small emittance, microbeam intensity is also small. However, the applications like a microbeam facility [7] require quite small intensities, down to 1-1000 particles/s. With some $10^9$-$10^{11}$ particles stored in the AGS ring, this gives enough room for constructing beams of very small emittances discussed above.

Let us take the example of AGS to estimate the achievable intensity of channeled microbeam. The beam circulating in the AGS ring has the size about ±5 mm before the extraction septum. The beam store is typically $10^9$ ions or $10^{11}$ protons. An area of 1 μm$^2$ would be hit by ~10 ions (or 1000 protons) in the time of a single turn in the ring (~1 μs at 1 GeV per nucleon); the hit rate is then ~$10^7$ ions/s per 1 μm$^2$.

The divergence of particles incident at crystal in periphery of the circulating beam, after crossing a stripping foil, is expected to be several times bigger than channeling acceptance. For particles trapped by a crystal or nanotube, the transmission factor would be 10 to 100% (e.g., Figs. 3-4) if channeled particles are bent a few mrad.

Microbeam intensity of $10^5$-$10^7$ ions/s appears even far greater than needed (though it is easily reduced by moving the crystal away from the core of the beam or misaligning it). One can put the question differently: how much a crystal can survive? The IHEP experience shows that crystals can channel up to ~$3 \cdot 10^{12}$ particles/s per cross-section of 0.5×5 mm$^2$ without cooling measures. This corresponds to $10^6/(s \cdot \mu m^2) = 1/(s \cdot nm^2)$. So, a micro-crystal structure can channel much more particles than needed, and even a nano-rope could do the job.

A lifetime of ~$5 \cdot 10^{20}$ proton irradiation per cm$^2$ as measured [14] for channeling crystal corresponds to $5 \cdot 10^{12}/\mu m^2$; this means over 100 years of operation of 1-$\mu m^2$ crystal with channeling of ~1000 protons/s, or one year for (20-nm)$^2$ nano-rope operating at 100 protons/s.

**Acknowledgements**. The authors thank K.A.Brown for discussions and R.P.Fliller for comments. This work was partially supported by INFN - Gruppo V, as NANO experiment, and by INTAS-CERN Grant No. 132-2000.

**References**

1. V.M. Biryukov, Yu.A. Chesnokov and V.I. Kotov, *"Crystal Channeling and its Application at High Energy Accelerators"* (Springer, Berlin, 1997)
2. M.B.H. Breese, *Nucl. Instr. and Meth. B* **132,** 540 (1997)
3. R.A.Carrigan, Jr., et al. *Phys. Rev. ST Accel. Beams AB* **1**, 022801 (1998). R.A.Carrigan, Jr., et al. *Phys. Rev. ST Accel. Beams AB* **5,** 043501 (2002)
4. A.G. Afonin, et al. *Phys. Rev. Lett.* **87**, 094802 (2001)
5. R.P. Fliller III, et al., EPAC 2002 Proceedings (Paris), p.200
6. T.W. Ebbesen, *Phys. Today*, **49**, 26 (1996). Z.Y. Wu, et al., *Appl. Phys. Lett.* **80**, 2973 (2002)
7. K.A. Brown, et al., EPAC 2002 Proceedings (Paris), p.554
8. P. Kleimann, J. Linnros, R. Juhasz. *Appl. Phys. Lett.* **79**, 1727 (2001)
9. V.V.Klimov and V.S.Letokhov, *Phys. Lett. A* **222**, 424 (1996)
10. L.G.Gevorgian, K.A.Ispirian, R.K.Ispirian. *JETP Lett.* **66**, 322 (1997)
11. N.K. Zhevago and V.I. Glebov, *Phys. Lett. A* **250**, 360 (1998)
12. V.M. Biryukov and S. Bellucci, *Phys. Lett. B* **542**, 111 (2002)
13. S. Bellucci, et al., *Nucl. Instr. and Meth. B* in press; [ArXiv:physics/0208081]
14. A. Baurichter et al., *Nucl. Instrum. Meth. B* **164-165**, 27 (2000)

**Figure captions**

**Figure 1**

Crystal with a micron-sized strip on the surface.

**Figure 2**

The number of channeled protons as a function of the nanotube curvature $pv/R$ for tubes of different diameters (bottom up: 2.2 and 1.1 nm). For comparison, also shown is the same function for Si(110) crystal (top curve).

**Figure 3**

The angular distribution of protons downstream of the bent nanotube, shown for two energies.

**Figure 4**

The angular distribution of $Fe^{+26}$ ions downstream of the bent nanotube, shown for two energies.

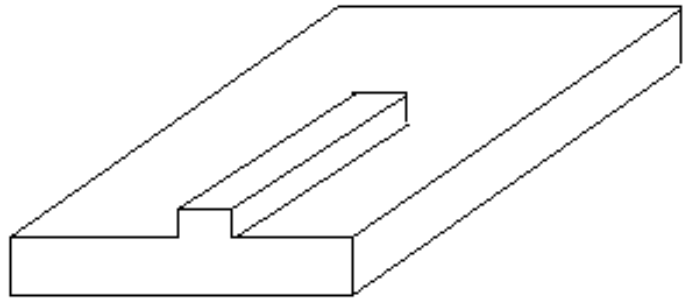

**Figure 1**

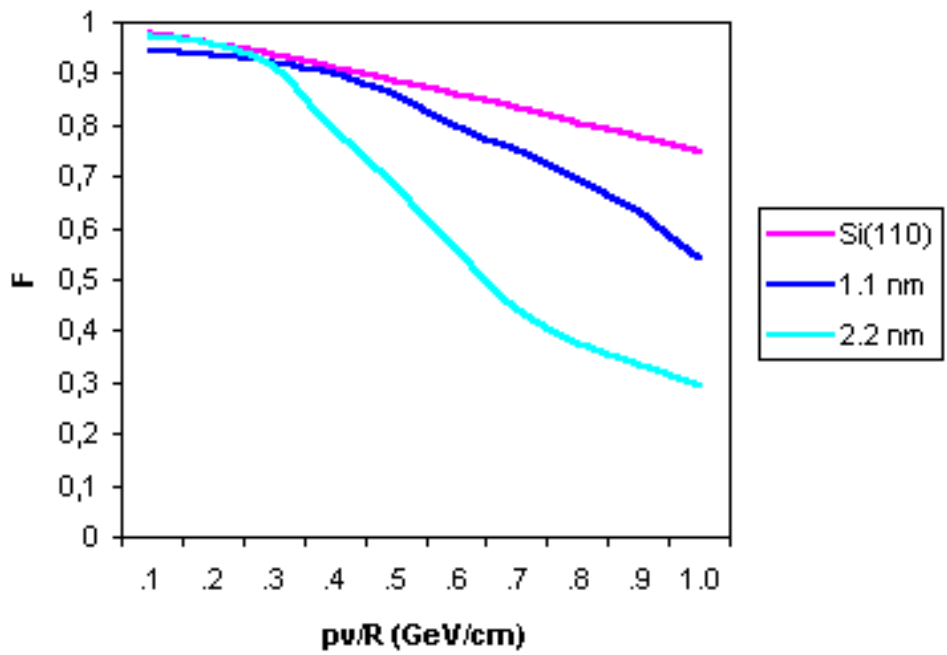

**Figure 2**

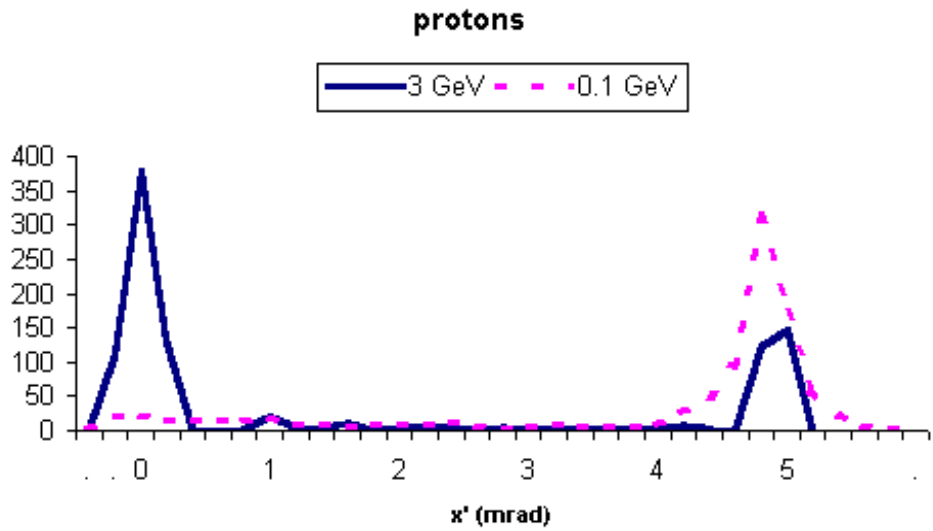

**Figure 3**

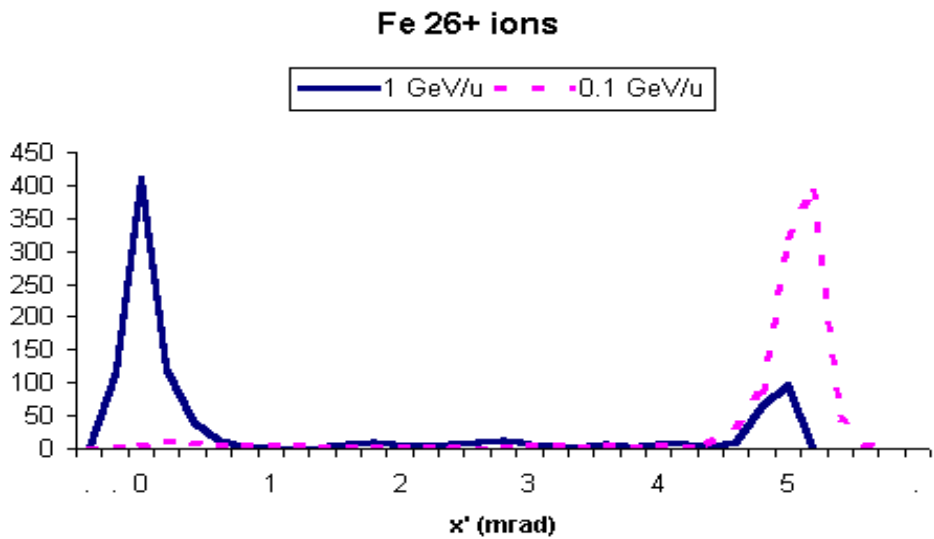

**Figure 4**